# Unconditional quantum teleportational advantage of single photons

Li-Chao Peng[1,2,3,*], Dian Wu[1,2,3,*], Xue-Mei Gu[1,2,3], Jian Qin[1,2,3], Ke-Mi Xu[1,2,3], Han-Sen Zhong[1,4], Hui Wang[1,2,3], Yu-Ming He[1,2,3], Ming-Chen Chen[1,2,3], Li Li[1,2,3], Nai-Le Liu[1,2,3], Chao-Yang Lu[1,2,3], Jian-Wei Pan[1,2,3]

[1]Hefei National Research Center for Physical Sciences at the Microscale and School of Physical Sciences, University of Science and Technology of China, Hefei, Anhui 230026, China

[2]CAS Centre for Excellence and Synergetic Innovation Centre in Quantum Information and Quantum Physics, University of Science and Technology of China, Hefei 230026, China

[3]Hefei National Laboratory, University of Science and Technology of China, Hefei 230088, China

[4]Shanghai Artificial Intelligence Laboratory, Shanghai 200030, China

**Abstract**

**Photon loss is the biggest enemy in quantum communications. Direct transmission typically results in a small fraction of photons surviving over long distances, hindering advanced applications such as loophole-free Bell tests and device-independent quantum key distribution. Quantum teleportation offers a "disembodied" way of transferring particles through a virtual channel—quantum entanglement—which, in principle, could overcome the photon transmission loss. Experimentally, however, no previous quantum teleportation of single photons has shown a higher survival probability than using direct transmission. To overcome this challenge, here we first propose and demonstrate an all-optical scheme for remote preparation of entangled photons. Through an effective 15-dB channel loss, we realize a heralding efficiency of 82% for event-ready entangled photons. Based on the prior distributed entanglement, we demonstrate that teleportation-based transmission efficiency offers a 2.95-fold enhancement, compared to using direct transmission through the same channel, therefore establishing an unconditional teleportational advantage.**

## Main text

Photons, the fast-flying qubits, are the natural candidate as carriers of quantum information to propagate over long distances in telecommunication optical fibers or free space[1]. However, the intrinsic absorption and scattering in these communication channels will cause photon loss that scales exponentially as a function of the channel distance. With optical fibers, typical channel loss is 15-20 dB per 100 km. Using Micius satellite, the space-to-ground channel loss is about 36 dB at a distance of 1000 km[1]. Due to quantum non-cloning theorem[2], the channel transmission loss is the biggest enemy for quantum communications.

Quantum teleportation[3] provides a conceptually new way to faithfully transfer arbitrary unknown quantum state to a remote location, without actually physically transmitting it through a lossy channel, and thus, can in principle overcome the transmission loss. Quantum teleportation has been realized in pure quantum photonic systems[4,5] as well as in matter qubits interconnected via photonic channels, such as atoms[6,7], trapped ions[8,9], solid-state spins[10,11], and optomechanical systems[12]. Yet, achieving a quantum teleportation of single photons to show an unconditional advantage compared to direct optical transmission remains challenging. This is primarily due to the requirement of pre-shared entanglement between distant nodes, which should be created in a heralded way and have high efficiency. However, this cannot be achieved using a brutal force distribution of entangled photon pairs (Fig.1a). For example, the heralding efficiencies of the entangled photons after satellite-based distribution was below $10^{-6}$ (ref. 13).

While entangled pairs can be created with near-unity collection efficiency locally[20], the challenge is to overcome the transmission loss and remotely generate entangled photons with near-unity heralding efficiency. Long-distance entanglement distribution with high heralding efficiency is central not only for faithfully realizing quantum teleportation but also for fundamental test of quantum mechanics[14-16], device-independent quantum key distribution[17], quantum metrologies[18], and distributed quantum computing[19].

There has been extensive research interest in this direction, both theoretically and experimentally. Entanglement swapping[22] in principle provides an elegant solution to generate high-efficiency entangled pairs at a distance. This scheme starts with two independent entangled pairs at locations A and B, which can be written as $|\phi^+\rangle_{1,2} =$ $(|H\rangle_1|H\rangle_2 + |V\rangle_1|V\rangle_2)/\sqrt{2}$ and $|\phi^+\rangle_{3,4} = (|H\rangle_3|H\rangle_4 + |V\rangle_3|V\rangle_4)/\sqrt{2}$, respectively, where $H$ and $V$ denote the horizontal and vertical polarization, and the subscripts denote their spatial modes. By projecting one qubit from each pair (say, 2 and 3) into a Bell state, the remaining two (1 and 4) becomes entangled without any direct interaction. While photon 2 and 3 travel through a certain channel distance, which thus do suffer from photon loss, it is important to note that photon 1 and 4 remain locally and are heralded to be entangled with high local collection efficiencies.

Yet, over 30 years after the beautiful idea[22], a pure photonic realization of faithful, free-propagating, and high-heralding-efficiency entanglement swapping remained elusive. The first demonstration[23] used entangled photons from spontaneous parametric down-conversion (SPDC)[24] which suffered both the double-pair emission and low collection efficiency (~1%). Because of the unavoidable double pair emission that contaminated two-photon events in the Bell-state projection, the experiment had to resort to post-selection, that is, destructive detection of the swapped photons 1 and 4 was necessary to validate the experimental protocol. Recent experiments[25,26] used two entangled photon pairs from the same quantum dot for entanglement swapping. The use of single quantum emitters suppressed double pair emission, however, the collection efficiency of single-emitter fluorescence was ~4%.

There were other theoretical schemes for overcoming photon loss in the quantum channel for quantum communication, such as heralded qubit amplifier[27,28] and sum-frequency generation[29], and the use of single quantum emitters[30]. However, these schemes have not yet proven feasible with current technologies. Another scheme[31], employing photon-number-discriminating detectors, only permitted the heralded creation of entangled photon pairs locally, thereby limiting its usability[32,33].

In this Article, we overcome the long-standing obstacle of photon transmission loss in quantum channels. We propose and realize a feasible protocol—using SPDC, linear optics, and commercial single-photon detectors—to realize unconditional entanglement swapping. At a 15-dB channel loss (an effective distance of 100 km), we create freely-propagating entangled photon pairs with high heralding efficiency of 82%. Next, based on the prior distributed two-photon entanglement, we demonstrate faithful quantum teleportation of a single photon, showing a 2.95-fold enhancement in photon survival efficiency compared to using direct transmission through the same lossy channel.

**Theoretical concept**

Our scheme for creating high-heralding-efficiency entangled photon pairs is shown in Fig. 1b. The key idea is to implement an unconditional quantum non-demolition (QND) measurement of the survival single photons after the long-distance travel. While post-selective (conditional) QND has been realized in the quantum teleportation of multiple degrees of freedom of a single photon[34], the challenge remained for the unconditional QND with imperfect SPDC source (with double-pair events) and conventional single-photon detectors.

Our protocol (Fig.1c) can be understood in the framework of a modified, unconditional entanglement swapping based on optimal SPDC sources. The necessary ingredients are three pairs of SPDC entangled photons which have near-unity indistinguishability and collection efficiency, simultaneously. Two pairs are created at the location of Alice ($|\psi\rangle_{1,2}$) and Bob ($|\psi\rangle_{5,6}$), respectively. The other ancillary pair ($|\psi\rangle_{3,4}$) is in the midpoint of the channel between Alice and Bob. The goal is that a four-photon (2-3-4-5) joint detection should herald a free-flying entangled photons at 1 and 6. The entangled photon pairs 1-2 and 5-6 must be designed to have near-unity collection efficiencies and meanwhile frequency uncorrelated, such that no photon loss would occur locally (at 1 and 6) from narrowband spectral filtering. In addition, the SPDC photons should preferably be at telecommunication wavelengths to minimize loss during long-distance transmission. Such a source became available in the 12-photon entanglement experiment[20].

The key to our protocol is to eliminate the erroneous events from multi-pair emission, the most notorious problem in SPDC. We define the probability of single-pair emission $\epsilon$ for $|\psi\rangle_{1,2}$ and $|\psi\rangle_{5,6}$, and $\gamma\epsilon$ for $|\psi\rangle_{3,4}$, where typically $\epsilon \sim 1\%$ and $\gamma \in [0,1]$. Photon 3 and photon 4 are distributed to Alice and Bob, respectively, over a distance with a total channel transmission efficiency $\eta$ ($\eta < 1$). Ideally, if there is one and only one photon in each spatial mode, a successful joint detection of photon 2-3-4-5 heralds an EPR pair $|\psi\rangle_{1,6}$, which occurs with a probability of $p_0 = \gamma\epsilon^3\eta/4$. The main noise that can erroneously trigger detectors 2, 3, 4 and 5 are from three main contributions: (1) double-pair emission of the first and third SPDC sources, occurring with a probability of $p_1 = \epsilon^4/16$. (2) double-pair emission from the midpoint SPDC source combined with single-pair emission from one of the outer SPDC, with a probability of $p_2 \simeq \gamma^2 \epsilon^3 \eta^2/8 + \gamma^2 \epsilon^3 \eta^{1.5}(1 - \eta^{0.5})/2$. (3) third-order emission from the midpoint SPDC sources, with a probability of $p_3 \simeq \gamma^3 \epsilon^3 \eta^2 (4 - 3\eta^{0.5})^2/64$. Higher-order terms are considered negligible due to the small value of $\epsilon$. Note that only the case (1) results in photons present in both modes 1 and 6. The two polarization-encoded Bell-state measurement (BSM) analyzers are designed at $|H\rangle/|V\rangle$ and $|+\rangle/|-\rangle$ basis, respectively, which effectively suppress the contamination of double-pair emission from the midpoint SPDC source (see supplemental information for details).

The estimated heralding efficiency and success rate (defined as $p_0/(p_0 + p_1)$) for generating the EPR state $|\psi\rangle_{5,6}$ are depicted in Fig. 1d. A near-optimal entangled pair, with both high heralding efficiency and state fidelity can be remotely generated between Alice and Bob, conditioned on $\epsilon < \gamma\eta \ll 1$. We adjust the emission probability $\gamma\epsilon$ for $|\psi\rangle_{3,4}$ in order to keep the total efficiency $\gamma\epsilon\eta$ at a constant value, which ensures a consistent six-photon count rate at different channel efficiencies $\eta$. As shown in Fig. 1d, the simulated heralding efficiency of our protocol (red curve) decreases in a concave-downward trend with increasing distance, in contrast to the exponential decrease for direct transmission (dashed curve). This demonstrates a significant advantage of our protocol for long-range entanglement distribution. Moreover, the success rate of our protocol remains consistently high—approximately 0.857—and shows almost no dependence on transmission distance. Interestingly, the parameter $\eta$, though typically considered a nuisance in most quantum communication tasks, becomes a useful asset in ensuring the effectiveness of our protocol.

### **Heralded entanglement generation**

In our experiment, we use three beam-like and frequency-uncorrelated SPDC sources to generate three EPR pairs at telecommunication wavelength. Each EPR pair is prepared in a Bell state $|\phi^+\rangle$, exhibiting a near-unity contrast of 0.99 in the $|+\rangle/|-\rangle$ basis, and an indistinguishability of 0.96 between independent single photons without narrowband filtering. These Bell states were then distributed to Alice and Bob via a quantum channel, where Bell-state measurements were performed. Specifically, BSM I discriminate between the Bell states $|\phi^+\rangle_{2,3}$ and $|\phi^-\rangle_{2,3}$, and BSM II distinguishes $|\phi^+\rangle_{4,5}$ and $|\psi^-\rangle_{4,5}$. Based on the joint measurement outcomes of BSM I and BSM II, we obtain the desired genuine Bell state $|\phi^+\rangle_{1,6}$ through a unitary transformation.

By adjusting the laser power to set $\epsilon \simeq 2\%$, we measured a two-photon count rate of about 600 kHz using commercial superconducting nanowire single-photon detectors (SNSPDs) with a system detection efficiency of about 75%. Two adjustable optical attenuators are employed to simulate the channel loss corresponding to the varying distances between Alice and Bob. The channel efficiency $\eta$ is set to five values: 1, 0.5, 0.18, 0.09 and 0.03. These values correspond to the equivalent transmission of photons through ultralow-loss fiber links (with attenuation below 0.15 dB/km at 1550 nm[35]) with lengths of 0 km, 20 km, 50 km, 70 km, and 100 km, respectively. Based on the triggered detections of four-photon (2-3-4-5), the quantum state shared between photon 1 and photon 6 can be expressed in Fock basis as $\hat{\rho} = \mathcal{A}_0 \hat{\rho}_{00} + \mathcal{A}_1(\hat{\rho}_{01} + \hat{\rho}_{10}) + \mathcal{A}_2 \hat{\rho}_{11}$, where $\hat{\rho}_{00}$ represents vacuum state, $\hat{\rho}_{01}$ and $\hat{\rho}_{10}$ correspond to states with a single-photon detection in either mode 1 or mode 6, and $\hat{\rho}_{11}$ denotes the two-photon entangled state. The coefficients $\mathcal{A}_i$ ($i$=0,1,2) denote the probabilities of each respective component and satisfy the normalization $\sum_i \mathcal{A}_i = 1$. Note that high-order terms are neglected due to their relatively low probability of occurrence.

The experimental heralded efficiency of entanglement generation can thus be extracted as $\tilde{\mathcal{A}}_2 = \tilde{c}_6/(\tilde{c}_4 \eta_1 \eta_6)$, where $\tilde{c}_6$ ($\tilde{c}_4$) is the six (four) fold coincidence rate registered by a time digital converter, $\eta_1$=61.1% and $\eta_6$=62.8% are the overall detection efficiencies for modes 1 and 6, respectively. The measured heralding efficiencies at different equivalent distances are described in Fig. 2a, indicating strong agreement between the experimental data and theoretical prediction. Remarkably, we achieve a heralded efficiency of (82.8±3.4) % for the entangled state $|\psi\rangle_{1,6}$ through a channel with an effective loss of 15-dB.

The fidelity of the reconstructed state $\hat{\rho}_{11}$ between photon 1 and photon 6 is evaluated using the formula $\mathcal{F}_{(1,6)} = \text{Tr}(\hat{\rho}_{11}|\phi^+\rangle\langle\phi^+|) = \text{Tr}[\hat{\rho}_{11}(\hat{I} + \hat{\sigma}_x \hat{\sigma}_x + \hat{\sigma}_z \hat{\sigma}_z - \hat{\sigma}_y \hat{\sigma}_y)]/4$, where $\hat{\sigma}_i$ ($i$=x, y, z) denotes standard Pauli operator. The experimentally measured fidelities for different equivalent distances are shown in Fig.2b. The results give $\mathcal{F}_0$ =0.773±0.021, $\mathcal{F}_{20}$ =0.828±0.016, $\mathcal{F}_{50}$ =0.765±0.021, $\mathcal{F}_{70}$ =0.759±0.021 and $\mathcal{F}_{100}$=0.751±0.020, here the subscript represents the equivalent distance in kilometers. These values, all significantly exceeding the classical limit of 0.5, clearly confirm the presence of genuine entanglement between photon 1 and photon 6. Despite some experimental imperfections, the measured entanglement fidelity and heralding efficiency remain considerably high even under fiber link losses characteristic of metropolitan-scale quantum network[36,37]. This robust performance highlights our protocol's potential for practical quantum information processing.

**Unconditional quantum teleportation**

To define quantum teleportational advantage, we consider what experimental strategies provide better transmission of a single photon over lossy and noisy quantum channels. A strategy is considered to offer a practical advantage if it achieves higher overall transmission efficiency while ensuring the fidelity of the transferred state not less than a benchmark fidelity $F_0$.

In the competing classical strategy, the direct transmission rate is $R_c = \eta$. Assuming a perfect classical transmission fidelity and a typical imperfect quantum teleportation fidelity, the classical strategy can be enhanced through optimal quantum state cloning[38], which involves generating $N$ local copies of the input quantum states. This improves the overall transmission efficiency by an $N$-fold factor—all $N$ photons are transmitted simultaneously—for a target output fidelity of $F_c = (2N + 1)/3N$.

By contrast, quantum teleportation provides a theoretical strategy capable of achieving unity transmission efficiency by leveraging prior shared entanglement between distant nodes. However, in previous experiments, the entanglement is locally generated and then distributed through lossy channels, which prohibits the effectiveness of quantum strategy. We show in Fig. 3 a lossless quantum teleportation scheme that effectively mitigates this problem.

By superimposing the photon to be teleported and one photon from the heralded EPR pair at a BSM and using classical communication, we can transfer quantum states with a practical efficiency $\tilde{R}_Q = \tilde{\mathcal{A}}_2 \tilde{\eta}_d^2 \tilde{\eta}_p / 2$ and fidelity of $\tilde{F}_Q$. Here $\tilde{\eta}_d$ represents the overall photon detection efficiency of each mode, $\tilde{\eta}_p$ is the preparation and detection efficiency of the to-be-teleported state and the factor of 1/2 comes from the success probability of the incomplete BSM with linear optics.

With an overall channel efficiency of 1% between the sender (*Alice*) and the receiver

(*Bob*) (see Implementation details in SI. B), the teleportation protocol proceeds as follows. First, a heralded EPR pair in the state $|\phi^+\rangle_{(1,6)}$ or $|\phi^-\rangle_{(1,6)}$ between photon 1 and 6 is preprared by selecting specific outcome from detectors 2-3-4-5. Alice then generates a heralded single photonic qubit from a SPDC pair (7,8) and performs a type I BSM on photon 7 and photon 6, which discriminates between $|\phi^+\rangle_{(6,7)}$ and $|\phi^-\rangle_{(6,7)}$. After that, Bob carries on the state measurements by using half- and quarter-wave plates and a polarizing beam splitter, with each optical mode detected by a SNSPD. To demonstrate the universality of the quantum teleportation process, Alice teleports six distinct input states chosen from a set of mutually unbiased bases, namely, $|H\rangle$, $|V\rangle$, $|+\rangle$, $|-\rangle$, $|R\rangle = (|H\rangle + i|V\rangle)/\sqrt{2}$ and $|L\rangle = (|H\rangle - i|V\rangle)/\sqrt{2}$.

The measured results are summarized in Fig. 4. The fidelity of the final teleported states $\tilde{\rho}$ is defined as $\tilde{F}_Q = \text{Tr}(|\psi\rangle\langle\psi|\tilde{\rho})$ and the quantum teleportation efficiency which indicates a success rate of quantum teleportation conditional on the heralding signals for single-photon (7) and EPR pairs (1,6)) is calculated as $\tilde{R}_{\text{Tel}} = \tilde{c}_8/\tilde{c}_5$, where $\tilde{c}_8$ and $\tilde{c}_5$ are eight-fold coincidence of eight-photons (1-8) and five-fold coincidence of photons 2-3-4-5-8, respectively. We obtain an average fidelity of $\bar{F}_{\text{Tel}} = 0.826 \pm 0.019$, exceeding the classical bound of 2/3 (ref.[39]), and indicating the non-classicality of our transferred states. Moreover, we measure an averaged teleportation efficiency of $\bar{R}_{\text{Tel}}$= (6.20±0.22) % over all six input states, surpassing the 1% efficiency limit achievable by direct brutal force state transfer.

For a transferred state fidelity of $F_0 = 0.826$, the optimal classical strategy utilizing optimal quantum state cloning achieves an improved transmission probability of 2.1%, assuming a perfect detection (see Fig. S12). Remarkably, even with imperfect single-photon detections, our experimental implementation already exhibits a 2.95-fold enhancement for realistic quantum information transfer, surpassing the classical bound by 18.6 standard deviations, unequivocally demonstrating the unconditional quantum advantage of our protocol. This advantage is expected to be enhanced through the use of actively multiplexed high-efficiency SNSPDs.

**Concluding remarks**

Our experiment fulfills the long-standing goal of free-flying photonic entanglement swapping and realizes unconditional quantum teleportation of a single photon with an advantage of a superior survival rate compared to direct transmission. The quantum teleportational advantage of single photons, marks another significant step forward that builds on previous demonstrations of quantum computational[40], metrological[41,42], and learning[43] advantages. In addition, our quantum teleportation protocol can be seen as a three-node quantum relay[44], in combination with quantum memories and heralded entanglement purification[21]. We anticipate the demonstrated quantum teleportational advantage will find applications in practical quantum networks, such as interconnecting quantum computers.

**Data availability**

All data are available from the corresponding author upon reasonable request.

**Acknowledgements**

This work was supported by the National Natural Science Foundation of China (Grant No. 12012422 and 62405025), the National Key R&D Program of China (Grant No. 2019YFA0308700), the Chinese Academy of Sciences, the Anhui Initiative in Quantum Information Technologies (Grant No. AHY060000), Innovation Program for Quantum Science and Technology (No. ZD0202010000) and the New Cornerstone Science Foundation. L.-C.P. acknowledges support from the China Postdoctoral Science Foundation (Grant No. 2024M754109).

**Competing interests:** The authors declare no competing interests.

## Figure Captions

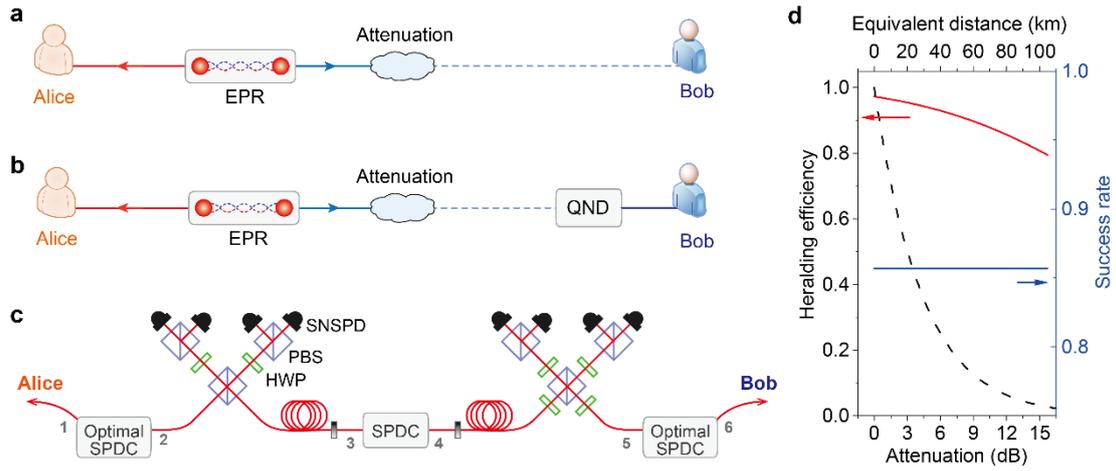

**Fig. 1 | Distribution of entangled photon pairs**. **a**, A direct entanglement distribution scheme, where Alice directly transfers one photon of the generated EPR pairs through a loss channel to Bob. In such a scheme, photon loss increases exponentially with propagation distance, thus severely limiting the applications of entanglement. **b**, Our scheme for heralded EPR pairs. A teleportation-based quantum non-demolition (QND) measurement is used to remove erroneous events caused by photon loss. **c**, Experimental scheme for high-efficiency heralded EPR pairs. Three EPR pairs (1-2, 3-4, 5-6) are generated via SPDC process. Two Bell-state measurement analyzers BSM I and BSM II are respectively performed on photons 2 and 3, and photons 4 and 5. In BSM I, photons are superimposed at polarization beam splitter (PBS) in $|H\rangle/|V\rangle$ basis and measured in $|+\rangle/|-\rangle$ basis. While in BSM II, photons are superimposed at circular-PBS in $|+\rangle/|-\rangle$ basis (which is constructed with a standard PBS and four half-wave plates (HWPs)) and measured in $|H\rangle/|V\rangle$ basis. All the photons are detected by superconducting nanowire single photon detectors. **d**, Theoretical analysis of the heralding efficiency (red line) and success rate (blue line) of the generated entangled photons as a function of channel efficiency. The dashed line represents the efficiency achievable using direct brute-force entanglement distribution.

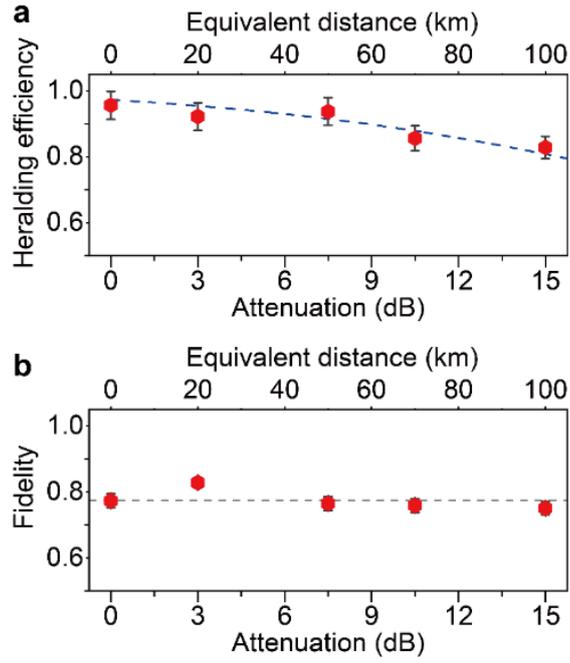

**Fig. 2 | Experimental results of the heralded entangled state $|\psi\rangle_{1,6}$. a,** Theoretical and experimental heralding efficiencies of $|\psi\rangle_{1,6}$ under different channel efficiencies. The red solid points are calculated from raw experimental data and the blue dashed line corresponds to theoretical prediction. **b**, Measured entanglement fidelity. The black dashed line depicts an average fidelity value of 0.775. Error bars represent one standard deviation, calculated according to Poissonian statistics of counts in the experiment.

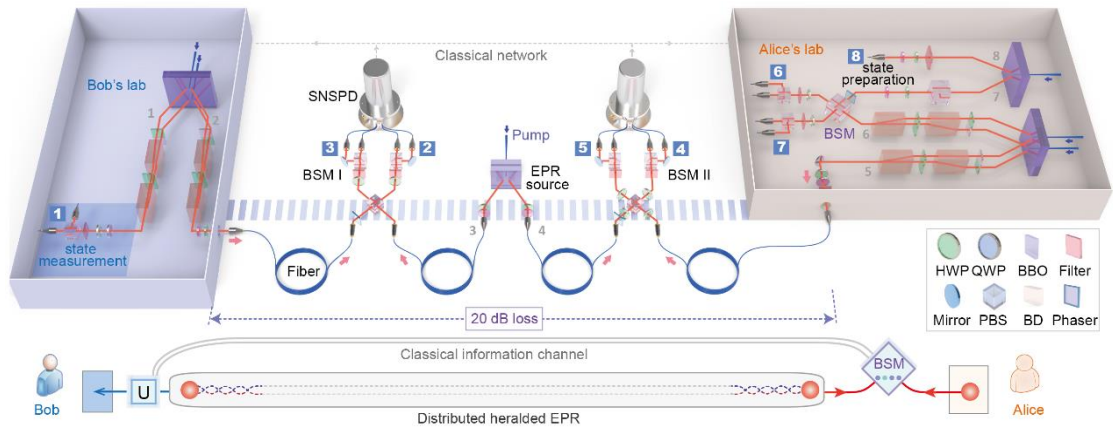

**Fig. 3 | Experimental setup and conceptual representation of unconditional quantum teleportation.** The top panel schematically illustrates our experimental setup, which consists of three core sections: Alice's laboratory for initial quantum state preparation and Bell-state measurement (BSM), an intermediate segment dedicated to the heralded distribution of entangled photon pairs, and Bob's laboratory for quantum state measurement. Each SPDC-based EPR source is pumped by a femtosecond pulsed laser operating at a wavelength of 775 nm with a repetition rate of 80 MHz. The generated down-conversion photons are spectrally filtered and centered at a wavelength of 1550 nm. Prior to each BSM, phase shifters and waveplates are employed to compensate for additional phase and polarization rotations introduced by the polarizing beam splitters (PBSs) and optical fibers. Channel attenuation is realized by incorporating precisely controlled optical fiber spools and calibrated optical attenuators to accurately simulate a range of transmission losses. The bottom panel provides a schematic diagram corresponding to the experimental setup, offering a simplified representation of the key components and their interconnections.

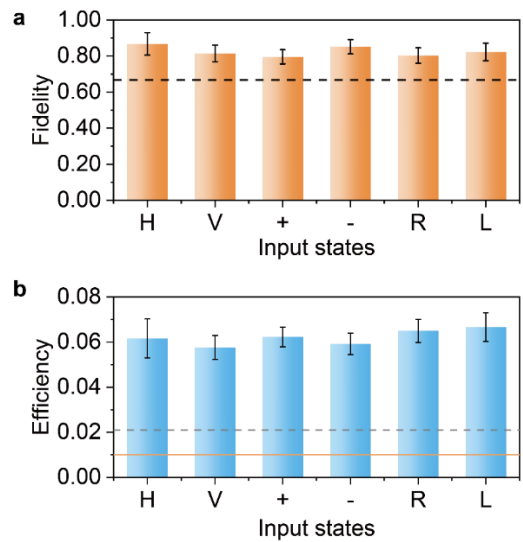

**Fig. 4 | Experimental results for quantum teleportation. a,** State fidelity measured for each input states, compared to the classical bound of 2/3 (dashed line). **b,** Teleportation efficiency for various input states, shown against the efficiency of direct classical transmission (orange line) and the classical efficiency bound (dashed gray line). Error bars represent one standard deviation.